\documentclass[9pt]{article}

\usepackage{epsfig}
\usepackage{latexsym}
\usepackage{amsfonts}
\usepackage{array}
\usepackage[english]{babel}
\usepackage{amsmath}
\usepackage{amssymb}
\usepackage{amsthm}
\usepackage{latexsym}
\usepackage{graphicx}

\def\bE{\begin{equation}}
\def\eE{\end{equation}}
\def\bEA{\begin{eqnarray}}
\def\eEA{\end{eqnarray}}
\def\bEAnn{\begin{eqnarray*}}
\def\eEAnn{\end{eqnarray*}}

\begin{document}
\title{\huge {\sc The Process of price formation and the skewness of asset returns}}
\author{Stefan Reimann\footnote{Contact address: sreimann@iew.unizh.ch}\\
Swiss Banking Institute\\ University of Zurich}
\maketitle

\begin{abstract}
\noindent
Distributions of assets returns exhibit a slight skewness. In this note we show that our model of endogenous price formation \cite{Reimann2006} creates an asymmetric return distribution if the price dynamics are a process in which consecutive trading periods are dependent from each other in the sense that opening prices equal closing prices of the former trading period. The corresponding parameter $\alpha$ is estimated from daily prices from 01/01/1999 - 12/31/2004 for 9 large indices. For the S\&P 500, the skewness distribution of all its constituting assets is also calculated. The skewness distribution due to our model is compared with the distribution of the empirical skewness values of the ingle assets. 
\end{abstract}

\section{Introduction}

The existence of stylized facts suggests that price trails of different financial markets might be regarded as different realizations of a more general stochastic system, called 'The financial market'. If so then the question is about the nature of this system. Since prices are macro-observables of a financial market, the model about price dynamics is defined on the macro level. Due to the set of assumptions used in its derivation, this model is an approximation in itself, see \cite{Reimann2006}. Three major properties estimated there were the distribution of (logarithmic) asset returns, the spectrum of Hurst exponents of their time series, and finally two entropy measures, the Renyi entropy and the Tsallis entropy. Although the model is a zero-order approximation, theoretical results agree with real data. First-order corrections concerning the set of assumptions made should improve these theoretical findings. An important feature the 'old' model was not able to capture is the (slight) skewness. In fact, if trading happens along a sequence of independent trading periods, i.e. opening prices are independent from closing price of the former trading period, the the distribution of returns is symmetric - in contrary to empirical data.

In fact the assumption that opening prices are independent from closing prices is economically unreasonable. If each period is one day at an exchange, then there is night between closing time and opening time. While on the one hand, many things can happen during the night time, it is not reasonable that the next opening price is independent from the price at the former closing time. In the contrary we assume that successive trading periods are dependent from each in that the opening price of each period equals the closing price of its former trading period.\\

To recapitulate the basic idea of our model: People go to the financial market to 'let their money work'. They do so by investing their money into assets. If the agent has capital $m_t$ to invest in asset $A$, he can buy $|A| = \frac{m_t}{X_t}$ units of this asset for its price $X_t$. A unit of this asset has an uncertain value $\delta_{t}$ some time later\footnote{Some therefore like to write $\delta_{t+1}$ for the corresponding adapted stochastic process.}. Hence at this time, the money $m_t$ invested in asset $A$ has value $
M_{t+1} = \frac{\delta_{t}}{X_t} \; m_t$.
If the agent is lucky, then at the expiration time $\frac{\delta}{x} > 1$, and his money $m$ has become more valuable by a factor
\bE \label{lambda}
\lambda_{t} \; = \frac{M_{t+1}}{m_t} =  \frac{\delta_{t}}{X_t}.
\eE
It is reasonable to assume that the agent wants to spend his money in an asset of which he expects that  $\frac{\delta_{t}}{X_t} > 1$. Thus, depending on his expectation about the future value of $\lambda_{t}$ the agents buys or sells this asset. Therefore, if the agent expects that  $\lambda_{t}> 1$, he will buy, otherwise he will sell. This causes an increase (decrease) of demand in this asset. Due to the increase (decrease) of demand, the price will rise (fall).  Thus the growth rate of the price process is some function of the recently expected gain $\lambda$. \\

The physical time interval $[0,t)$ is divided into intervals marking trading periods $T_\tau = [\tau,\tau+1)$, where $[ \tau $ can be regarded as its opening time and $\tau+1)$ as its closing  time. Considering daily prices, it might be suggestive to think that a trading period lasts for one day, from the opening of the exchange to its closing. Two periods are then separated by night time, in which no trade but a lot can happen. 
$$
[ \; 0 , \; t \; )  \; = \;  [ \; 0, \; 1) \; \star [ \; 1, \; 2) \; \star \hdots \; \star [ \; t-1, \; t)
$$
Denoting the price by a function $X$, we say that $X_t := X_{[t}$ is the opening price of period $T_t$, while $X'_t:=X_{t)}$ is the closing price of this period. Trading periods can be dependent or independent in the sense that, in the first case, the closing price $X'_t$ equals the opening price of the next period $X_{t+1}$. If periods are independent, the price evolution is an ensemble property of a 1-period model defined for a single period which arbitrary initial prices $X$, so that the price process becomes 'quasi - stationary'. Otherwise price evolution is described by a multiplicative process in time given by 
\bE\label{X}
X_{t+1} \; = \;  \varphi(\lambda_{t}) \: X_t.
\eE
where the growth rate $\varphi(\lambda)$ is assumed to be an increasing function of $\lambda$. 
The model and its analysis considered here is based on the following four assumptions:
\begin{enumerate}
\item The financial market contains only 1 asset;
\item Trading periods are {\em dependent}; 
\item Payoffs are uniformly distributed within a fixed finite interval;
\item The growth rate $\varphi$ is a power law with a constant scaling exponent $\alpha$, i.e.
\bE\label{phi}
\varphi(\lambda) \; = \; \lambda^\alpha \; = \; X^{-\alpha}, \qquad \alpha>0.
\eE
\end{enumerate} 
As seen, compared with the model in in \cite{Reimann2006}, the only difference is in assumption 2, i.e. trading periods are no longer independent from each other. As we will see this will induce asymmetry in the system. 

\subsection{$\alpha$ and asymmetry}

$\alpha$ acts as a zooming factor in that, if $\alpha < 1$, the growth rate gives highest weight to price fluctuations on a small level, while if $\alpha > 1$, price fluctuations on a high level are more important. This is since $
\partial \phi(X) \sim X^{\alpha-1}
$ and hence for $Y > X$, 
$$
\partial \phi(Y) > \partial \phi(X) \qquad \mbox{iff} \qquad \alpha > 1.
$$
Second, if $\alpha<1$, then negative returns are more probable than positive returns, and {\em vice versa}: Recall that, given two successive prices $X',X$, the return $Z = \ln \frac{X}{X}$ yields
$
Z = \alpha \; \ln \frac{\delta}{X}
$ 
Consequently, $\partial_\alpha Z_t = \frac{1}{\alpha} \; Z$ which is solved by
$$
Z(\alpha) = Z \: \ln(\alpha) 
$$
Therefore, if $\alpha<1$, negative returns $Z(\alpha) < 0$ are more probable than positive returns. Taking both arguments together this gives the following picture: Talking about expectations, the parameter $\alpha$ can be regarded as a 'preference parameter' in the sense that if $\alpha < 1$ the agent puts more weight on low level price fluctuations. On the other hand, if $\alpha>1$ the agent puts more weight on high level price fluctuations. On the other hand $\alpha$ can be regarded as an asymmetry parameter: If $\alpha \not= 1$, there exists an asymmetry in the model, which then is reflected in the skewness of returns. In this sense, skewness - negative for $\alpha<1$ and positive for $\alpha>1$, is a consequence of differently weighting price fluctuations that happen on different levels.

\section{A formal exercise}

This argumentation can be made precise by considering price evolution as a process, i.e. assuming that $X_{t+1} = X'_t$, in which using equations \ref{lambda}, \ref{X}, and \ref{phi}, consecutive prices are related by 
\bE\label{model}
X_{t+1} \; = \; \delta_t ^\alpha \quad X_t^{1-\alpha}.
\eE
Logarithmic prices $\zeta_t = \ln \: X_t$ then satisfy the difference equation
$
\zeta_t := \alpha \ln \delta_t + (1-\alpha)\;\zeta_{t-1} 
$, whose generating function - for $\zeta_0 = 0$ -  yields
\bE
F_\alpha(s) \; = \; \alpha \sum_{t\ge 1}\frac{ \ln \delta_t  \: s^t}{1-s+\alpha s}
\eE
The coefficients $c_\alpha(t)$ of its Taylor expansion in $s=0$ obey
$
c_\alpha(t) = \zeta_t
$ and hence $Z_\alpha(t) = \zeta_t - \zeta_{t-1} = \ln \frac{X_t}{X_{t-1}}$ is obtained from
\bE\label{Zalpha}
Z_\alpha(t) = c_\alpha(t)- c_\alpha(t-1).
\eE
If $\alpha = 1 - a$, $|a|  \ll 1$,
expansion of $Z_\alpha(t)$ in equation \ref{Zalpha} around $\alpha = 1$ up to first order in $\alpha$ then gives
\bEA
Z_{\alpha}(t) &=& \ln \bigg( \delta^{1-a}_t \; \delta^{2a-1}_{t-1} \; \delta^{-a}_{t-2}\bigg) \; + \; {\cal O}(a^2)
\eEA
$Z_\alpha$ is the sum of the following random variables $Y_j$ derived from $\delta_{t-j} \sim {\cal U}(0,1)$, $j=0,1,2$ with probabilities $f_{Y_j}(z)$ respectively
$$
\begin{array}{ccl lcl}
Y_0 &=& (1-a) \: \ln \delta_{t} &, \quad f_{Y_0}(z) &=& c_0 \; e^\frac{z}{1-a} \: I_{(-\infty,0)}\\
Y_1 &=& (2a-1) \: \ln \delta_{t-1} &, \quad f_{Y_1}(z) &=& c_1 \; e^\frac{z}{2a-1} \: I_{(0, \infty)}\\
Y_2 &=& -a \: \ln \delta_{t-2} &, \quad f_{Y_2}(z) &=& c_2 \; e^\frac{z}{-a} \: I_{(0, \infty)}
\end{array}
$$
where normalization constants yield $c_0 = \frac{1}{1-a}, c_1 = \frac{1}{1-2a}, c_2 = \frac{1}{a}$.
Since the $Y_j$ are independent, the probability density of the compound variable $Z_\alpha$ is the convolution of the densities of the compound variables, i.e.
\bEA
f_Z (z) & = & \left\{
	\begin{array}{ll}
	c_1 \: c_2 \; (f_{Y_1} \star f_{Y_2})(z) & z> 0\\
	c_0 \; f_{Y_0}(z) & z \le 0.
	\end{array}
				\right.
\eEA
Thus, up to a normalization for $|Z| \gg 0$, the distribution is given by
\bE\label{asymdistr}
f_Z (z) \; = \; \left\{
	\begin{array}{lr}
	 \frac{1}{1-3a} \; e^{-\frac{z}{1-2a}} & z\ge 0\\
	\frac{1}{1-a} \; e^\frac{z}{1-a}& z \le 0.
	\end{array}
	\right.,
\eE 
Therefore, in a semi-logarithmic plot we see a tent - with an exponential correction for small $z$ - according to 
$$
	\ln f_{Z_\alpha}(z) \; \sim \; \left\{
	\begin{array}{lc}
		- \frac{z}{1-2a} & z> 0\\
		+ \frac{z}{1-a} & z \le 0
	\end{array}
	\right.,
$$
\begin{figure}[h]
  \setlength{\unitlength}{1mm} 
\begin{center}

  \begin{minipage}[t]{60mm}
  \hskip 1.5cm
	\begin{picture}(60,40)
		\thinlines
		\put(0,5){\framebox(38,30)}
		\put(-16,18){$\ln f_{Z_\alpha}(z)$}
		\put(25,25){$-\frac{1}{2\alpha-1}$}
		\put(8,25){$\frac{1}{\alpha}$}
		\put(18,30){\line(1,-1){16}}
		\put(16,30){\line(-1,-2){7}}
		\put(19,0){$0$}
		\put(-3,30){$0$}
		\put(33,0){$z$}
	\end{picture}
\caption{Positive skewness of the return distribution for $\alpha>1$}
  \end{minipage} \hskip 5mm
    \begin{minipage}[t]{60mm}
    \hskip 1.5 cm
	  \begin{picture}(60,40)
		\thinlines
		\put(0,5){\framebox(38,30)}
		\put(-16,18){$\ln f_{Z_\alpha}(z)$}
		\put(23,30){\line(1,-2){6}}
		\put(21,30){\line(-1,-1){16}}
		\put(21,0){$0$}
		\put(-3,30){$0$}
		\put(33,0){$z$}
	\end{picture}
	\caption{Negative skewness of the return distribution when $\alpha<1$}
  \end{minipage} 
\end{center}
\end{figure}

For $\alpha=1$ the distribution is symmetric. 
\bE\label{symdistr}
\ln f_{Z_1}(z) \; = \; -\ln \: 2\; - \; | z | .
\eE
If $\alpha< 1$ $(a>0)$, positive returns are less probable than in the symmetric case, while if $\alpha> 1$, $(a<0)$, positive returns are more probable. Hence a positive $a$, i.e. $\alpha<1$, relates to negative skewness while $\alpha>1$ corresponds to positive skewness. This is in contratst to the case where trading periods all independent, since there the distribution is symmetric for all $\alpha$, see \cite{Reimann2006}
\bE
\ln f_{Z_\iota}(z) \; = \;- \ln (2\alpha) \: - \frac{| z |}{\alpha} \qquad \alpha > 0.
\eE

\subsection{Estimating $\alpha$ from data}

Given our model, one would like to estimate the parameter $\alpha$ from data. The obvious method is to compare the distributions of positive and negative returns: Since $\alpha_+ = 1-2\:\alpha$ and $\alpha_- = \alpha$, this would give
$$
\alpha = 1 + (\alpha_+ - \alpha_-).
$$
A better way to approximately estimate $\alpha$ uses the moments of the distribution. For small $\alpha$ we consider the case that $z \gg \frac{a(1-2a)}{1-a}$. Then, according to equation \ref{asymdistr}, the distribution yields
$$
f_Z (z) \; = \; c' \; \left\{
	\begin{array}{lr}
	 \frac{1}{1-3a} \; e^{-\frac{z}{1-2a}}  & z> 0\\
	\frac{1}{1-a} \; e^\frac{z}{1-a}& z \le 0.
	\end{array}
	\right.,
$$
whose normalization constant $c' = \frac{3a-1}{5a-2} \approx 1/2$ for small $a$, so that $\int f_Z(z) dz = 1 + {\cal O}(a^2)$. Its raw moments $\mu'_n$ of the distribution follow from its characteristic function 
$$
\Phi(t) = {\cal F}(f_Z)(t) = \frac{1}{2(1-3a)} \; \frac{2 \bigg( 1+i \:a(1-a) \; t\bigg)}{ \bigg(1+i \: t \:(1-a)\bigg) \: \bigg(1- i\:t(1 - 2a)\bigg)} 
$$
according to $\mu'_n = (-i)^n\Phi^{(n)}(0)$, from which central moments $\mu_n$ are obtained as their binomial transforms $\mu_n \; = \; \sum_{k=0}^n { n \choose k} (-1)^{n-k} \mu'_k  \mu'^{n-k}_1$. The skewness of the distribution then is
\bE\label{skewness}
\gamma_1 = \frac{\mu_3}{(\mu_2)^\frac{3}{2}} \; \approx \; - \frac{3}{\sqrt{2}} \; a \; + {\cal O}(a^2).
\eE
Therefore $\alpha$ can be estimated from data by
$$
\alpha \;=\; 1 + \frac{\sqrt{2}}{3} \; \gamma_1^{emp}
$$
The accuracy of the model can be estimated by inserting $a = 1-\alpha$ back into equation \ref{skewness} and to compare the true (empirical) skewness $\gamma^{emp}_1$ with the skewness $\gamma^{theor}_1$ estimated on the basis of our model. Both are compared by $\Delta = \frac{\gamma^{emp}_1 - \gamma^{theo}_1}{\gamma^{emp}_1}$, see table \ref{Alphas}. 
\begin{table}[h]
\begin{center}
\begin{tabular}{|l|ccc|c|}
\hline
index & $\alpha$ & emp $\gamma_1$ & theor $\gamma_1$ & $\Delta$ \\ \hline\hline
DAX 30 						 & 0.949053 & -0.108074 & -0.115446 & -0.068  \\ 
DOW JONES				 & 0.938115 & -0.131277 & -0.142329 & -0.084 \\ 
FRANCE CAC 40				 & 1.003512 &  0.007450 & 0.007417  & 0.004 \\ 
FTSE 100					 & 0.993641 & -0.013490 & -0.013598 & -0.008 \\ 
HangSeng					 & 1.126184 &  0.267676 & 0.231176  & 0.136 \\ 
NASDAQ 100				 & 1.152170 &  0.322801 & 0.271138  & 0.160 \\ 
NIKKEI 500					 & 1.052247 &  0.110832 & 0.104040  & 0.061 \\ 
S\&P 500				        & 0.993413 & -0.013972 & -0.014088 & -0.008 \\ 
SWISS SMI 					 & 0.971440 & -0.060585 & -0.062830 & -0.037 \\ 
\hline
\end{tabular}
\caption{\label{Alphas} For each index, we considered daily data from $01/01/1990$ to $12/31/2004$ provided by {\sc Thompson Datastream}}
\end{center}
\end{table}
As seen from the table, indices are skewed, some have positive skewness, some have negative skewness. Except the {\sc Hang Seng} and the {\sc Nasdaq}, the estimated $\alpha$ gives a reasonable skewness $\gamma_1^{theor}$ compared with the true skewness $\gamma_1^{emp}$. Both deviate from each other by at most 7\% only. Recalling that our analysis is of first-order only is quite satisfactory. What is wrong with the estimates for the {\sc Hang Seng} and the {\sc Nasdaq}? It will turn out from the next section, that these indices contain asset with rather high skewness and hence the first order approximation becomes poor.

\section{The profile of an index}  
An index $K$ is constituted by a number of selected assets $i_K$. Therefore an index should be characterized by the distribution of skewness values of its constituting assets rather than by a single skewness value. For the range see table \ref{Gammas}.
\begin{table}[h]
\begin{center}
\begin{tabular}{|l|cc |}
\hline
index & $\min_i \gamma_1^{emp}(i)$ & $\max_i \gamma_1^{emp}(i)$\\ \hline\hline
DAX 30 						 & 0.0050   & 2.0110  \\ 
DOW JONES				 & -1.9049   & 0.3856 \\ 
FRANCE CAC 40				 & -0.4862    &1.8777 \\ 
FTSE 100					 & -1.8597    &3.3351 \\ 
HangSeng					 &-0.1724   &\underbar{49.0180} \\ 
NASDAQ 100				 & -0.5906   &\underbar{20.5612}\\ 
NIKKEI 225					 & -0.5431    &1.7967\\ 
S\&P 500				        & -5.4587    &5.3549\\ 
SWISS SMI 					 &-1.5023    &0.8412 \\ 
\hline
\end{tabular}
\caption{\label{Gammas} For each index, the interval $[\min_i \gamma_1^{emp}(i), \max_i \gamma_1^{emp}(i)]$ is given, $\gamma^{emp(i_K)}$ are distributed in. }
\end{center}
\end{table}
As seen from Table \ref{Gammas}, all indices contain assets whose skewness' varies from positive to negative values.  There is actually one exception which is the {\sc DAX}.  It should be further noted that the {\sc Hang Seng} and the {\sc Nasdaq} contain assets with very high skewness. For them our low order approximation becomes poor, of course.\\

\begin{figure}[h]
  \setlength{\unitlength}{1cm} 
\begin{center}
  \begin{minipage}[t]{6cm}
  \begin{picture}(6.6,7)
      \includegraphics[width=60mm]{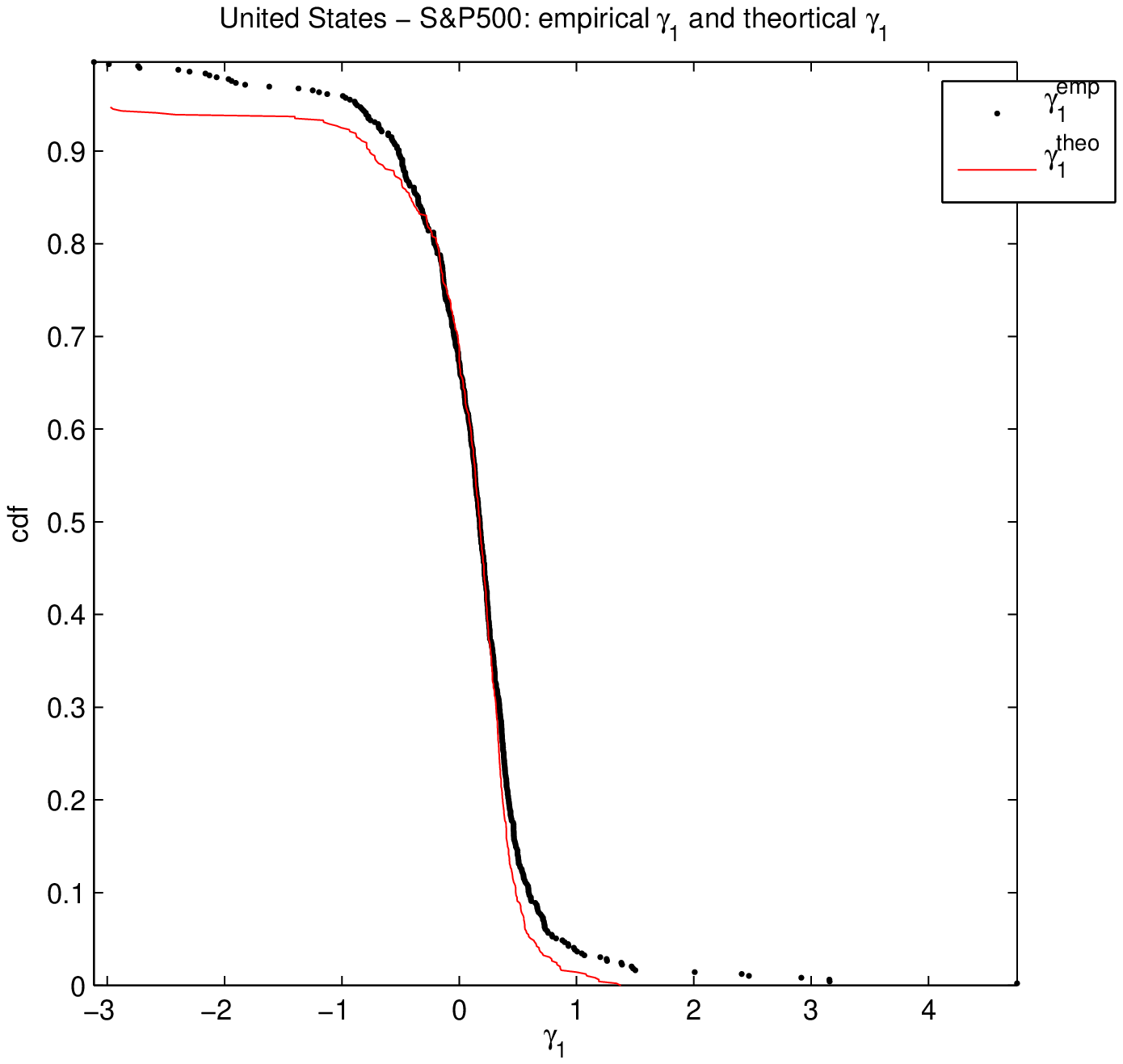}
  \end{picture}
  \end{minipage} \hskip 0.5 cm
  \begin{minipage}[t]{6cm}
  \begin{picture}(6.6,7)
    \includegraphics[width=60mm]{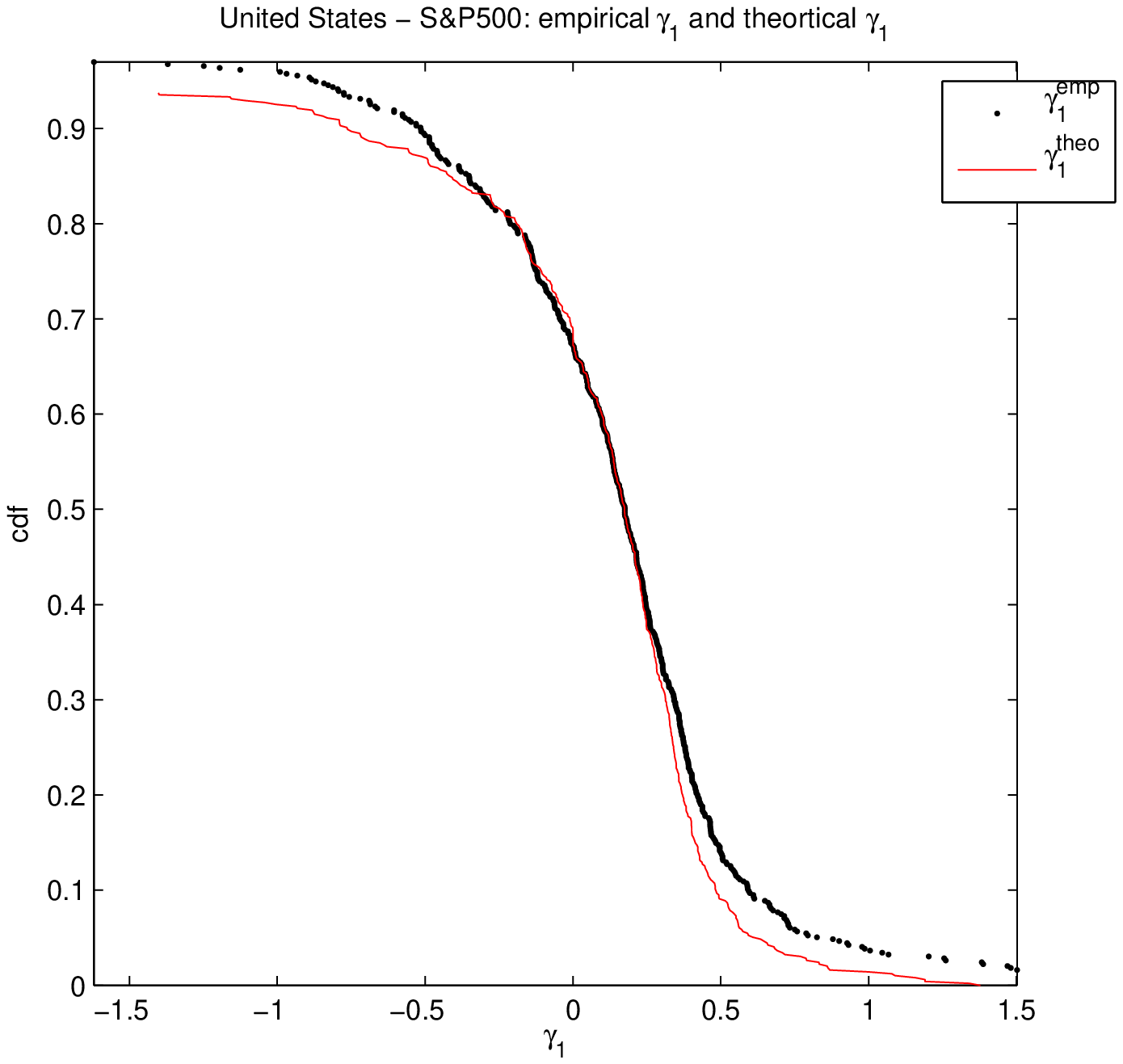}
  \end{picture}
  \end{minipage}
\end{center}
\caption{\label{sp500} \small Cumulative density functions $c^{emp}$ (black dots) and $c^{theor}$ (red line) for the {\sc S\&P 500}. The right picture is the central part of the right one. } 
\end{figure}

\noindent We determined the empirical skewness $\gamma_1^{emp}(i_K)$ for each asset $i_K$ as well as its estimated skewness $\gamma_1^{theor}(i_K)$. The cumulative density function of index $K$ then is
$$
c_K(\gamma) := {\mathbb P}[\gamma_1(i_K) \ge \gamma].
$$
In the following we consider the S\&P 500 and all its constituting assets. Cumulative density function for $\gamma^{theo}_1(i)$ and $\gamma^{emp}_1(i)$ are shown in Figure \ref{sp500}: The black line displays the empirical cdf, while the dashed line is the theoretical cdf. The estimated $\gamma_1^{theor}$ can only be a reasonable approximation for $\alpha \approx 1$. Therefore the right picture shows the central part of the diagram seen left. 

\section{Conclusion}
The aim is understand basic principles of price formation on a financial market in terms of a simple model. While the mechanism proposed might seem to be plausible, only the comparison with real data can judge about the feasibility of this model. \\
 
In \cite{Reimann2006} we proposed a simple model for endogenous price formation on the macro level. This model and its analysis was based on a series of assumptions. This made the model a Zero-order approximation. Comparison with real data showed that the proposed model did already a good job. On the other hand, its limitations were clearly seen, as outlined in \cite{Reimann2006}. The logically next step was to modify the made assumptions step for step to see whether the model is enhanced. One important feature,  the previous model was not able to reproduce, was the (slight) skewness of the distribution of asset returns seen in empirical data. Our previous analysis was concerned with the setting that all trading periods are independent from each other. In this case the return distribution was always symmetric. The analysis was based on the assumption that trading periods were independent from each other. The assumption that successive trading periods are independent from each other was not reasonable: Opening prices are not independent from closing price of the former trading period. Thus we substituted our initial assumption that trading periods are independent by the assumption that the opening price equals the closing price of the former period. Thus price evolution now became a process in time. The analysis in this note shows that, if the trading periods are dependent from each other, the return distribution is skew. The only parameter $\alpha$ of our model can be regarded as an {\em asymmetry parameter}. If $\alpha<1$, the distribution is negatively skewed, while if $\alpha>1$, it is positively skewed.  Only for $\alpha=1$, the distribution is symmetric.\\

In economic terms $\alpha$ can be given other interpretation. Following \ref{Reimann2006}, the parameter $\alpha$ might de understood as the long-term averaged {\em liquidity} of an asset. The discussion in section 1.1 additionally suggests an interpretation in term of {\em preferences}. The decision to buy or to sell the asset depends on the agents 'believe' about the future growth rate of the value of the asset. This depends on both, the recent price level and the fluctuations on this level. As the discussion in section 1.1 shows, if $\alpha<1$, then the agent puts more weight on price fluctuations on a low level, while otherwise, price fluctuations on a high level are more important. This induces an asymmetry due to deviations of $\alpha$ from $1$. This interpretation implies that the skewness of empirical asset returns are due to an asymmetric preference in the agents decision.\\ 

The next question was how much of the observed skewness of a empirical return distribution can be 'explained' as a consequence of considering trading as process? We therefore estimated the skewness parameter $\alpha$ from data and calculated the skewness that our model would generate given this parameter. This 'theoretical' skewness was then compared with the empirical skewness in the data. It turned out that a huge amount of empirical skewness is quite well described by our model, in which the price dynamics are a process!

\end{document}